\documentclass[twocolumn,pra,aps,showpacs]{revtex4}

\usepackage{graphicx}

\begin{document}
\title{Formation of shock waves in a Bose-Einstein condensate}
\author{Bogdan Damski}
\affiliation{
Instytut Fizyki, Uniwersytet
Jagiello\'nski,
ul. Reymonta 4, PL-30-059 Krak\'ow, Poland \\ and \\
Institut f\"ur Theoretische Physik, Universit\"at Hannover, D-30167
Hannover, Germany
}

\begin{abstract}
We consider propagation of density wave packets in a  Bose-Einstein condensate.
We show that the shape of initially broad, laser-induced, density perturbation 
changes in the course of free time evolution
so that a shock wave front finally forms. 
Our results are well beyond predictions of commonly used zero-amplitude
approach, so they can be useful in extraction of a speed of sound
from experimental data. We discuss a simple experimental setup for shock
propagation and point out possible limitations 
of the mean-field approach for  description of shock phenomena in a BEC.

\end{abstract}

\pacs{03.75.Kk,47.35.+i}
\maketitle

\section{Introduction}
\label{1}

Spectacular creation of a Bose-Einstein condensate (BEC) in 
atomic traps \cite{cornell}
has opened  unique possibilities for controlled studies of 
ultracold bosonic gases \cite{anglin}. 
One of the basic problems in 
physics of 
a BEC is  how density perturbations propagate
throughout a bosonic  cloud. There has been a big  interest  in this problem
and both an experiment \cite{kettfale} and  theoretical studies
\cite{kavoulakis,quasi} have been performed. These theoretical calculations 
have  been done, however, 
for infinitesimally small perturbations, which dynamics hardly
exhibits  nonlinear effects leading to formation of
shock waves. This paper is devoted to theoretical studies of
a shock wave formation in  homogeneous and harmonically trapped BEC. 


Shock waves have been widely investigated
in different physical systems \cite{whitham}.
For example, we find them in a  gas bubble driven
acoustically  \cite{sono}, in  a photonic crystal \cite{photo} and 
even in  mathematical models of traffic flow \cite{whitham}.
Of particular importance are studies of shock waves 
in classical compressible gases, where they 
appear whenever  large enough
density perturbation propagates \cite{landau,somerfeld}. Contrary to the case of classical gases,
shock waves in ultracold "quantum" atomic gases still require 
both theoretical and experimental investigations even though some 
results have been already obtained. 
The propagation of steep density perturbations in a BEC has been 
experimentally investigated in 
\cite{hau}; the theory of shocks in Fermi (Tonks) 
gas has been proposed in \cite{bodzio}; 
an approximate theoretical description of shock waves 
in a BEC has been worked out in \cite{kraenkel}. It is also worth to mention
that shocks have been studied in physical systems governed by nonlinear
Schr\"odinger equation \cite{ruski,kamchatnov}.
These studies can be related to investigations of shock waves in a BEC 
due to formal similarity between the nonlinear Schr\"odinger
equation and mean-field equations of a BEC. 
Nevertheless, it is important to realize that dynamics of shock waves even 
in weakly
interacting Bose gas, may differ from predictions of 
the mean-field approximation, as will be proposed in Sec. II.

The mean-field description of a BEC is provided by 
the Gross-Pitaevskii equation \cite{dalfovo}, which in the hydrodynamical variables and dimensionless form 
 \cite{units} reads
\begin{eqnarray}
 \label{h}
 \frac{\partial \rho}{\partial t} + \frac{\partial}{\partial x}\left(v\rho\right)
 =0, \nonumber \\ 
 \frac{\partial v}{\partial t} + \frac{\partial}{\partial x}\left(\frac{1}{2} v^2
 +V\right)+\frac{\partial}{\partial x}
 \left(g\rho-\frac{1}{2}\frac{\partial_x^2 \sqrt{\rho}}{\sqrt{\rho}}\right) = 0.
 \end{eqnarray}
First equation is a continuity equation, while the second one 
is similar to 
 the Euler equation from classical hydrodynamics. Naturally, $\rho$ is 
 atomic
 density and   $v$ is a velocity field.
The atomic density in (\ref{h})  
satisfies $\int dx \, \rho=1$, a constant $g>0$ is proportional to the 
strength of two-body repulsive interactions between atoms. The term 
$\propto \partial_x^2\sqrt{\rho}/\sqrt{\rho}$ is called quantum pressure (QP).
It is comparable to the $g\rho$ term only if the spatial scale of density
variations is of the order of the healing length
$\xi(\rho)$.
Simple estimation  on the basis of (\ref{h}) shows that 
$\xi(\rho)=1/\sqrt{2g\rho}$. 

In the following we assume that a BEC is initially in a state
having a Gaussian-like density wave packet created by  adiabatically
focused laser beam on an atomic cloud.
Alternatively, an atomic cloud can be cooled
in a presence of a laser beam, as proposed in \cite{kettfale}.
We assume that the width of a  laser beam 
is large compared to the system's healing length,
so that the density perturbation is broad and the QP term, at least initially,
is negligible. We are interested in  dynamics of such a 
laser-induced perturbation after
 abrupt laser turn off. We show that it splits into two wave packets
propagating in  opposite directions. These wave packets  undergo self-steepening
dynamics. Finally,  two shock wave fronts are  formed.

\section{1D homogeneous systems}

First we consider the case of a one-dimensional (1D) 
BEC in a periodic box 
having  boundaries at $x=\pm l$. 
There is initially a laser beam 
producing the potential $V=u_0\exp(-x^2/2\sigma^2)$ with $\sigma \ll l$ and $u_0$
being controlled by  intensity and detuning of a laser beam. 
The initial density of atoms takes form
\begin{equation}
\label{dens0}
\rho(x,0)=\rho_0+ 2 \eta \rho_0 e^{-x^2/2\sigma^2},
\end{equation}
where $\rho_0=1/(2l+2\eta\sqrt{2\pi}\sigma)$ provides
 normalization of 
(\ref{dens0}).
We assume that the system is in the ground state which implies
$v(x,0)=0$, and that the QP term
is negligible: $\sigma\gg \xi(\rho_0)$. It means that we are  
in the so-called Thomas-Fermi (TF) regime \cite{dalfovo}. 
Under such conditions $2\eta\rho_0=-u_0/g$.

 Let us first inspect what happens in a low amplitude limit. The analysis
 can be done in the framework of a Bogoliubov approach or within a
 hydrodynamical
 approach. The first possibility was explored for ultracold fermions
 in \cite{bodzio}. Here we follow the second one  
 to give the reader more insight into the problem. 

\begin{figure}
\centering
\includegraphics[scale=0.3,clip=true]{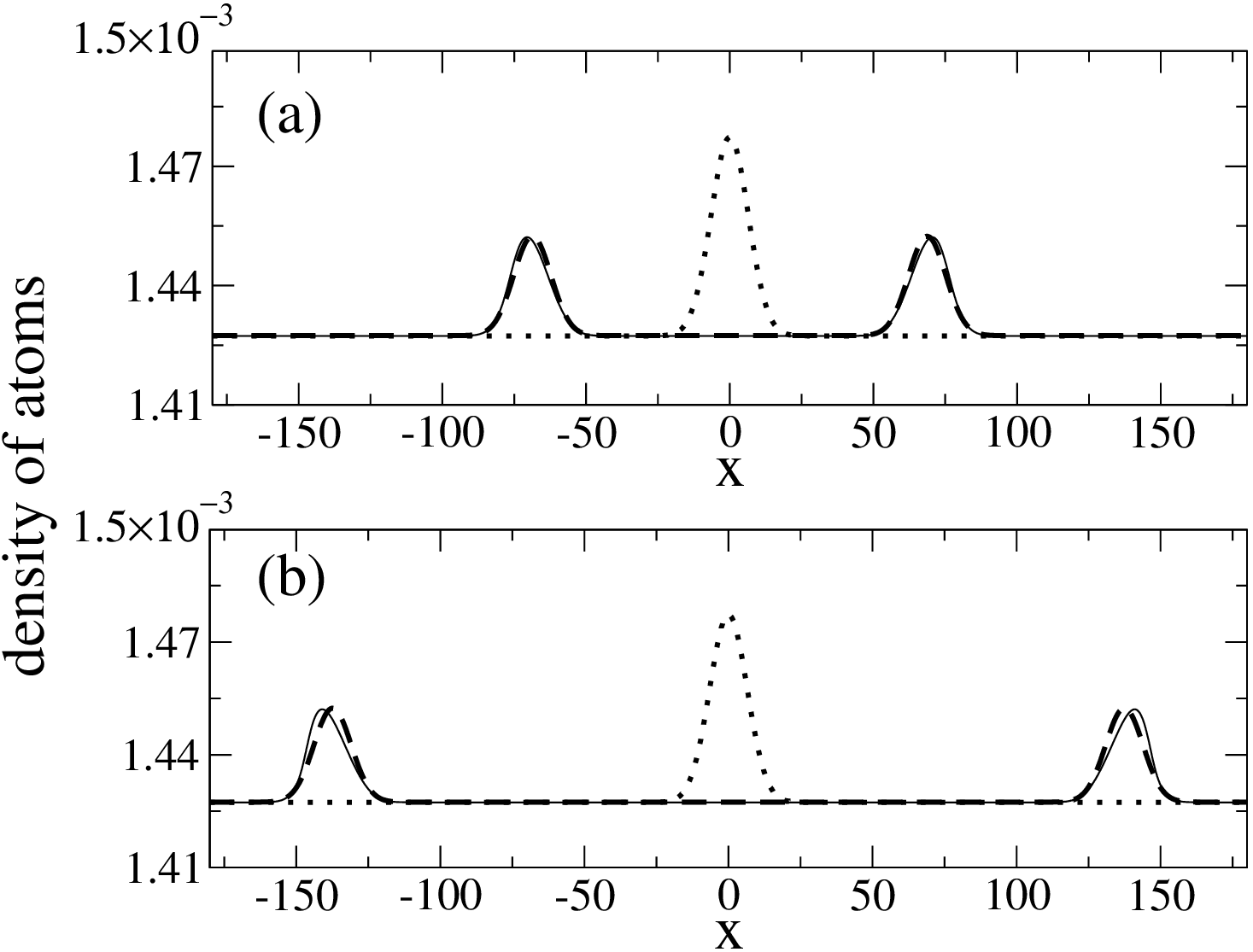}
\includegraphics[scale=0.3,clip=true]{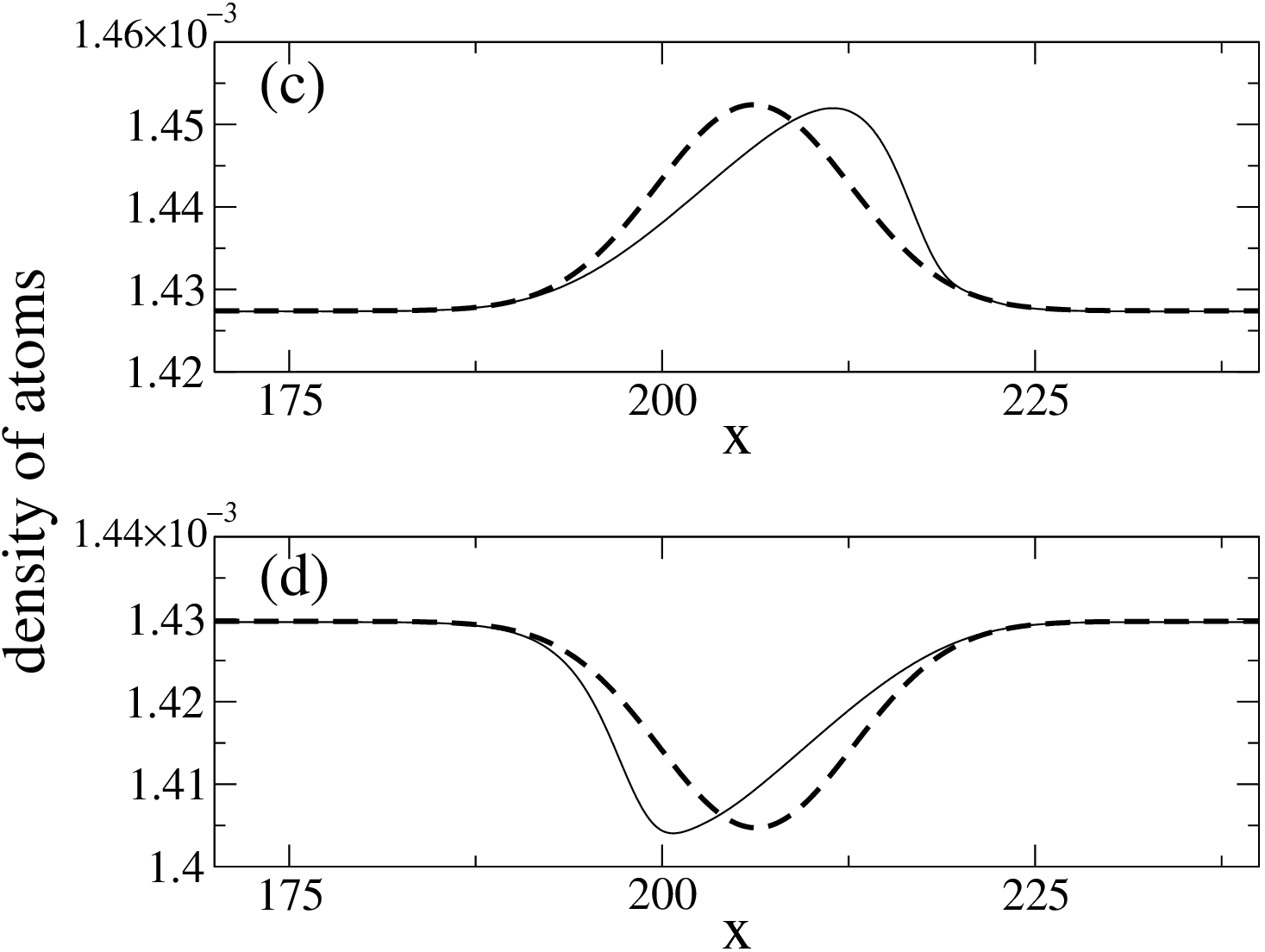}
\caption{Density of atoms during free time evolution.
Dotted line -- an initial density profile, 
dashed line -- the prediction (\ref{drho}), solid line -- a numerical solution of
(\ref{h}) with the QP term. Plot (a) [(b)]:  $\eta=0.0175$ and 
$t=21$ [$t=42$].
Plot (c) [(d)]: the right-moving wave packet for $\eta=0.0175$ [$\eta=-0.0175$]
at $t=63$. Other parameters: $g=7.5\cdot10^3$, $\sigma=6.5$, $l=350$.
}
\label{fig1}
\end{figure}

We express $\rho(x,t)$ as $\rho_0+\delta\rho(x,t)$ with 
$|\delta\rho(x,t)/\rho_0|\ll 1$. Having in mind that the velocity field 
$v(x,t)$ is of the same order as $\delta\rho(x,t)$, a linearization of 
 (\ref{h}) without the QP term gives
\begin{eqnarray}
\label{wave}
\left(\frac{\partial^2}{\partial t^2} -
c_0^2 \frac{\partial^2}{\partial x^2}\right)\phi(x,t)= 0,  \nonumber \\
v(x,t)= \frac{\partial\phi}{\partial x} \ \ , \ \  
\delta\rho(x,t)=-\frac{1}{g}\frac{\partial\phi}{\partial t},
\end{eqnarray}
where $c_0=\sqrt{g\rho_0}$ is a speed of sound wave propagating on  density
$\rho_0$ \cite{sound}.
Straightforward calculation determines the solution of (\ref{wave})
that satisfies 
initial conditions $\delta\rho(x,0)=2 \eta \rho_0\exp(-x^2/2\sigma^2)$
and $v(x,0)= 0$:
$$
\phi(x,t)=\frac{\eta\rho_0\sigma}{\sqrt{2\pi}}\int dk e^{-k^2\sigma^2/2}g[\sin(kx_+)-\sin(kx_-)]/kc_0,
$$ 
with $x_\pm=x\mp c_0t$. Using the explicit form of $\phi(x,t)$ and (\ref{wave})
we find 
\begin{equation}
\label{drho}
\delta\rho(x,t)= \eta\rho_0 \left[e^{-(x-c_0t)^2/2\sigma^2} + 
                              e^{-(x+c_0t)^2/2\sigma^2}\right].
\end{equation}
Therefore, the initial perturbation splits into two pieces 
traveling in opposite directions with a speed of sound
during free time evolution (Fig. \ref{fig1}a-b). 
Such a process is impossible in a system which dynamics is
governed by a single-particle Schr\"odinger equation. It is a nice 
interaction-induced phenomenon in a BEC. 
Interestingly, 
a similar effect can be
observed in Fermi (Tonks) clouds \cite{bodzio}.
It is also easy to see from (\ref{wave})
that the splitting phenomenon happens in exactly the same way  
regardless of the form of an initial shape.

Let us also remark that, since we consider
particles in a periodic box, all the equations should be strictly speaking
written in an appropriate Fourier basis. For $\sigma\ll l$ and 
$c_0t<l$ the discrepancy between obtained results and exact ones is hardly
visible. 

Even though the solution (\ref{drho}) was derived for  small perturbations
($|\eta|\ll 1$) it can be found numerically that the splitting of an initial
wave packet  into two oppositely moving pieces takes place for
 $|\eta|\sim1$ as well. Quantitative comparison of
 (\ref{drho}) to
 exact numerical results  reveals discrepancies
 growing  in the course of time evolution for any $|\eta|$.
An exact solution for $\eta>0$ ($\eta<0$) moves faster (slower) and changes
its shape so that the front (back) impulse's edge self-steepens -- Fig.
\ref{fig1}c-d. 

To explain that
behavior we have found, following \cite{landau}, 
the  exact traveling wave  
solution of Eqs. (\ref{h}) without the QP term:
\begin{equation}
\label{nonlin}
\rho=f(x-(a_0\pm3\sqrt{g\rho})\,t) \ \ \ , \ \ \
v=a_0\pm2\sqrt{g\rho},
\end{equation}
where $f$ (an arbitrary function) and $a_0$ (a constant) can be determined
from initial conditions.
The sign $\pm$ determines a direction
of propagation. 
The result (\ref{nonlin}) is valid as long as a spatial scale of density 
perturbations is  larger 
than the healing length $\xi$. 
Solutions of the form (\ref{nonlin}) were previously 
worked out in various physical systems, see e. g. 
\cite{kraenkel,ruski,kamchatnov}.

Nonlinearity of  (\ref{h}) prevents usage of the
superposition principle to built a solution, that contains both  left and
right-moving perturbations, out of (\ref{nonlin}). Fortunately, the 
problem can be simplified, since after  separation perturbations
move independently. 

We pro\-pose to consider only the right-moving (r) part 
and to take as an initial density profile 
$\protect{\rho_r(x,0)= \rho_0+ \rho_0\eta e^{-x^2/2\sigma^2}}$
(compare (\ref{dens0})). 
	Such an approximation is exact for small perturbations
and turns out to be  good for  large ones -- see discussion of numerical
simulations presented below.
The constant $a_0$ 
is found to be $-2\sqrt{g\rho_0}$ from a requirement that initially 
the
velocity field far from the perturbation is zero: $v(x\to\pm l,0)=0$.
This assumption leads to the solution
\begin{equation}
\label{an1}
\rho_r(x,t)=\rho_0+ \rho_0\eta\exp(-[x-\sqrt{g}
(-2\sqrt{\rho_0}+3\sqrt{\rho_r})t]^2/2\sigma^2).
\end{equation}
Although (\ref{an1}) is  in an implicit form, 
all the quantities characterizing a shock wave
formation can be analytically
extracted from it.

Both  amplitude and velocity of impulse's maximum are constant during time
evolution.
The amplitude equals $\rho_0(1+\eta)$, while the velocity  becomes
$\protect{{\cal{V}}(\eta)=\sqrt{g\rho_0}{\cal A}(\eta)}$ where
\begin{equation}
\label{a}
{\cal A}(\eta)=3\sqrt{1+\eta}-2.
\end{equation} 
For small perturbations
 we get ${\cal{V}}(\eta)=\sqrt{g\rho_0}(1+3\eta/2)$.
Taking $\eta=0$ we find a well known expression for a sound velocity:
$\sqrt{g\rho_0}$ \cite{sound}. 
These results show explicitly that bright perturbations ($\eta>0$) 
move faster than dark ones ($\eta<0$). A typical measurement of a sound
velocity relies on observation of propagation of density perturbations 
on a  cloud \cite{kettfale}. If such perturbations are not small enough
our results indicate that the $\eta$ dependent term in ${\cal{V}}(\eta)$
has to be considered in the experimental
determination of a sound
velocity.

For clarity of discussion we will concentrate now on {\it bright perturbations}.
A speed of impulse's maximum is bigger than a speed of
its tails. As a result an impulse self-steepens in the direction of
propagation so that formation of a  shock wave front takes place. Finally,
$\partial_x\rho_r(x,t_s)=-\infty$ at $x=x_s$ leading to the well  known 
wave breaking phenomenon \cite{whitham}.

\begin{figure}
\centering
\includegraphics[scale=0.3,clip=true]{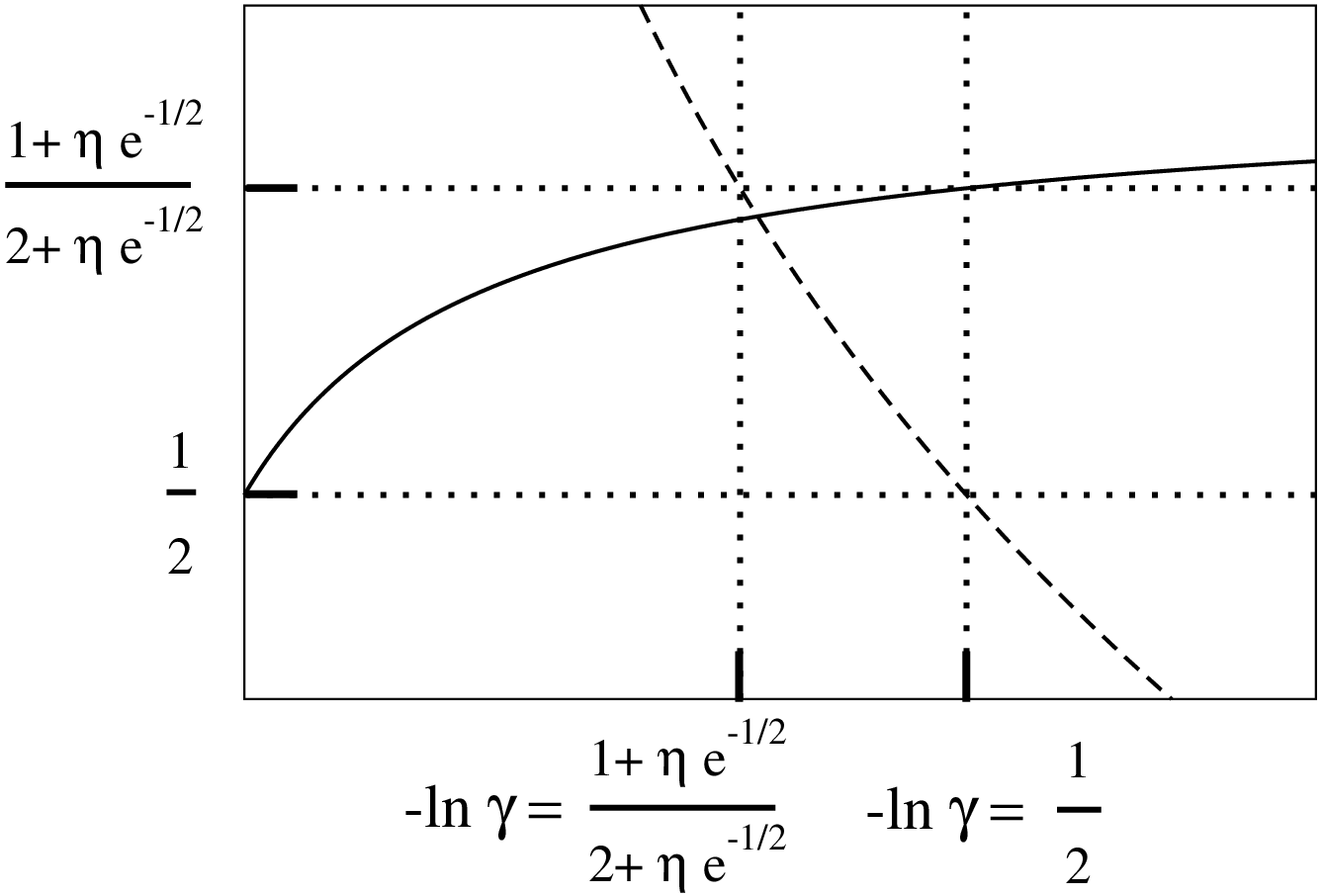}
\caption{Schematic plot of functions: $-\ln\gamma$ (dashed line) and
$(1+\eta\gamma)/(2+\eta\gamma)$ (solid line) vs. $\gamma$.
Dotted lines define bounds leading to inequalities (\ref{an6}).
}
\label{schemat}
\end{figure}

Time required for such a process ($t_s$) can be estimated as follows: 
the impulse's half-width ($\approx2\sigma$) is a 
difference in distance traveled by lower and upper  impulse's parts 
until the shock appears: 
$\protect{({\cal{V}}(\eta)-{\cal{V}}(0))t_s \approx 2\sigma}$.
It gives 
\begin{equation} 
\label{tshock}
t_s\approx\frac{2\sigma}{3(\sqrt{1+\eta}-1)\sqrt{g\rho_0}}.
\end{equation}
Quantitatively time of shock creation and shocks' position can be determined 
from a set of equations \cite{landau}
\begin{equation}
\label{an2}
\frac{\partial x(\rho_r)}{\partial \rho_r}= 0 \ \ \ ,  \ \ \
\frac{\partial^2 x(\rho_r)}{\partial \rho_r^2}= 0.
\end{equation}
The first one says that $\partial_x\rho_r$ is infinite, while the
second one assures  uniqueness of the $\rho_r(x,t_s)$ function.
Assuming $\eta>0$ we find that
\begin{equation}
\label{an3}
x(\rho_r)=\sqrt{g}(-2\sqrt{\rho_0}+3\sqrt{\rho_r})\,t + 
\sqrt{-2\sigma^2\ln[(\rho_r/\rho_0-1)/\eta]]},
\end{equation}
which substituted into (\ref{an2}) leads to 
\begin{equation}
\label{an4}
t_s=\frac{\sqrt{2}\sigma}{3\sqrt{g}}\frac{\sqrt{\rho_s+\rho_0}}{\rho_s-\rho_0},
\end{equation}
where $\rho_s$, the density at which $\partial_x\rho_r=-\infty$,
satisfies 
\begin{equation}
\label{an5}
-\ln\left[(\rho_s/\rho_0-1)/\eta\right]=\frac{\rho_s}{\rho_s+\rho_0}. 
\end{equation}
It seems to be impossible to solve analytically that equation.
Fortunately,  its numerical treatment is straightforward. It is possible,
however, to prove analytically that the solution of (\ref{an5}) exists and
satisfies the relation
\begin{equation}
\label{an6}
\rho_0 + \rho_0\eta e^{-{\cal B}(\eta)}\le\rho_s \le \rho_0 + \rho_0\eta e^{-1/2},
\end{equation}
where 
$$
{\cal B}(\eta)=\frac{1+\eta e^{-1/2}}{2+\eta e^{-1/2}}.$$
For $\eta\to0$ lower and upper bounds coincide. One can derive inequalities
(\ref{an6}) by
means of a graphical method -- Fig. \ref{schemat}. 
Let us introduce a variable $0\le\gamma\le1$ such that
$\rho_s=\rho_0(1+\eta\gamma)$.
Eq.  (\ref{an5}) now takes the form:
$-\ln\gamma=(1+\eta\gamma)/(2+\eta\gamma)$. 
The RHS of this equation
increases monotonically with $\gamma$ from $1/2$ to $\protect{(1+\eta)/(2+\eta)}$.
Conversely, the LHS decreases monotonically with $\gamma$ from
$+\infty$ to $0$.
Therefore there must be $\gamma$ at which the RHS  equals the LHS.
For $\gamma>\exp(-1/2)$ the LHS becomes smaller than $1/2$. 
 It means that $\gamma\le\exp(-1/2)$ in the considered problem.
Therefore, the RHS takes the highest value for $\gamma=\exp(-1/2)$,
so it is bounded from above by ${\cal B}(\eta)$. As a result, we arrive at
a conclusion that the solution exists and 
$1/2\le-\ln\gamma\le{\cal B}(\eta)$
leading to inequalities (\ref{an6}).

On the basis of (\ref{an4}) and (\ref{an6}) we have found bounds for  
time of shock  formation
\begin{eqnarray}
\label{an7}
t_s &\le& \frac{\sqrt{2}\sigma e^{{\cal B}(\eta)}}{3\eta\sqrt{g\rho_0}}\sqrt{2+\eta e^{-{\cal B}(\eta)}},
\nonumber \\
 t_s &\ge&\frac{\sqrt{2}\sigma e^{1/2} }{3\eta\sqrt{g\rho_0}}\sqrt{2+\eta e^{-1/2}}. 
\end{eqnarray}
In the $\eta\to0$ limit both bounds equal 
$2\sigma e^{1/2}/3\eta\sqrt{g\rho_0}$. 
From inequalities (\ref{an7}) 
we see that the shock  appears if a wave packet evolves long enough. 
This observation  together with the explicit expression for a speed of
propagation of  density maximum is  a central result of this paper.  
Considerations (\ref{wave})-(\ref{drho}) fail to predict such behavior since they are valid in
the $\eta\to0$ limit where $t_s\to\infty$. As mentioned above these
calculations have been  done for bright wave packets $(\eta>0)$.
Extension to 
dark impulses is straightforward. We would like also to point out that 
the solution (\ref{an1}) is uniquely defined 
in  a whole space up to the  time  of shock
creation, i.e. for $t\le t_s$ -- see 
Fig. \ref{large}b and Ref. \cite{drazin}.
Later on, the solution (\ref{an1})  in a 
shock wave front region becomes multi-valued, compare
\cite{kraenkel,ruski,kamchatnov}.

Approximating the position of a shock wave front 
by position of impulse's maximum at $t=t_s$: $x_s={\cal V}(\eta)t_s$, we get
\begin{eqnarray}
\label{xs}
x_s &\le& \frac{\sqrt{2}\sigma e^{{\cal B}(\eta)}}{3\eta}
\sqrt{2+\eta e^{-{\cal B}(\eta)}} {\cal A}(\eta),
 \nonumber \\
x_s &\ge& \frac{\sqrt{2}\sigma e^{1/2}}{3\eta}
\sqrt{2+\eta e^{-1/2}} {\cal A}(\eta).
\end{eqnarray}
As $\eta\to0$ we have $x_s\to2\sigma e^{1/2}/3\eta$.

Let us compare numerical simulations of hydrodynamical equations
with the QP term to above derived analytical predictions.
We must be aware that analytical calculations have been performed
without taking the QP term into account. 
Initially a spatial scale of density variations
is roughly $\sigma$, which is much greater than the healing length $\xi$.
As a result the QP term is negligible at the beginning of time evolution. 
As shocks form up density changes occur on smaller
and smaller length scales being finally of the order of $\xi$. Therefore,
before appearance of shocks quantum pressure has to start to play a role
leading to discrepancy between analytical predictions and numerical
simulations.

\begin{figure}
\centering
\includegraphics[scale=0.3,clip=true]{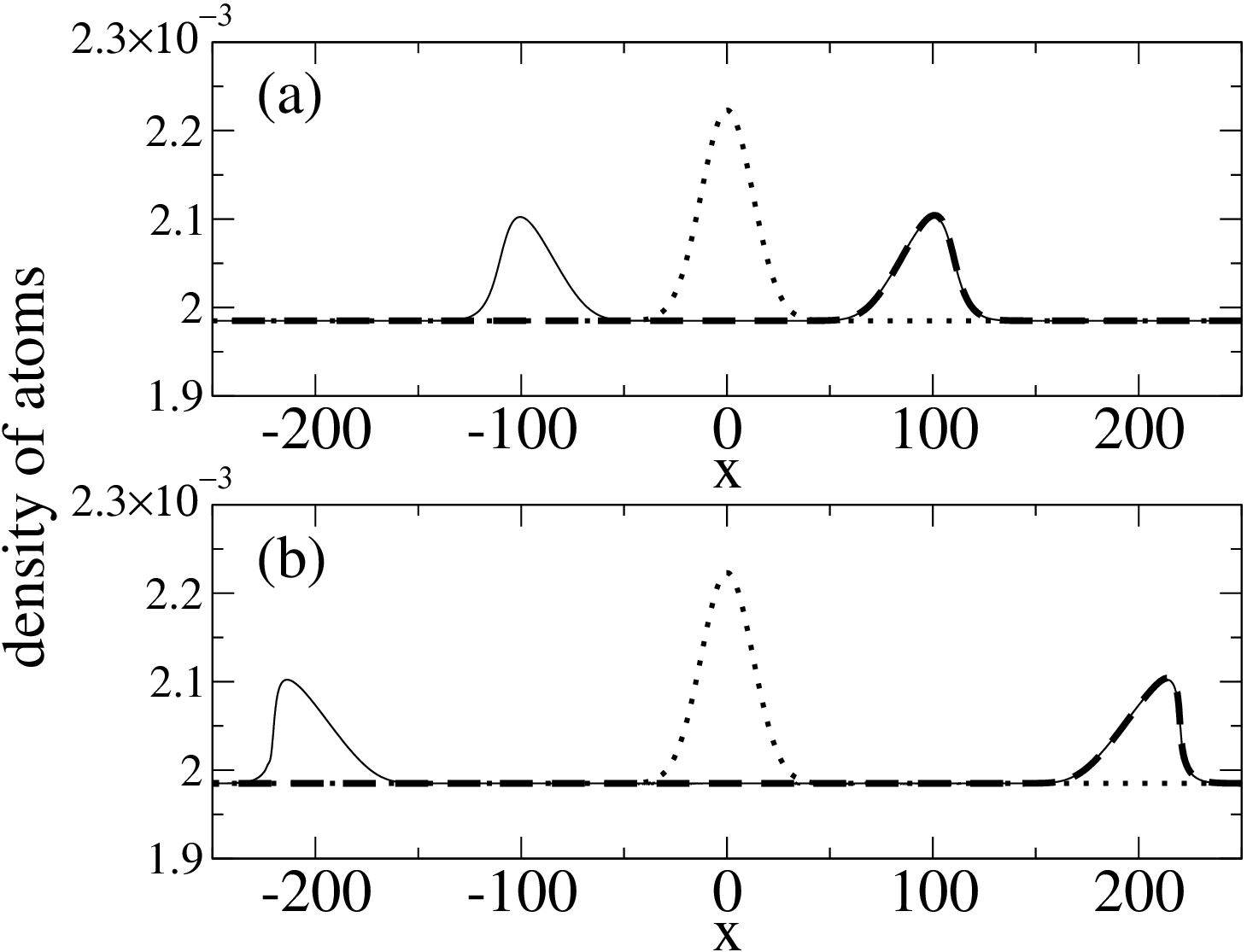}
\caption{Density of atoms during free time evolution. 
Dotted line -- an initial density of atoms, dashed line -- the 
analytical solution 
(\ref{an1}), solid line -- a numerical solution of Eq. (\ref{h}) with the QP
term.
Plot (a) [(b)] is for
$t=24$ [$t=51$]. Other parameters: $\eta=0.06$, 
$t_s=60.4$, $x_s=253.8$, $g=7.5\cdot10^3$, $\sigma=12.5$, $l=250$.
}
\label{medium}
\end{figure}

\begin{figure}
\centering
\includegraphics[scale=0.3,clip=true]{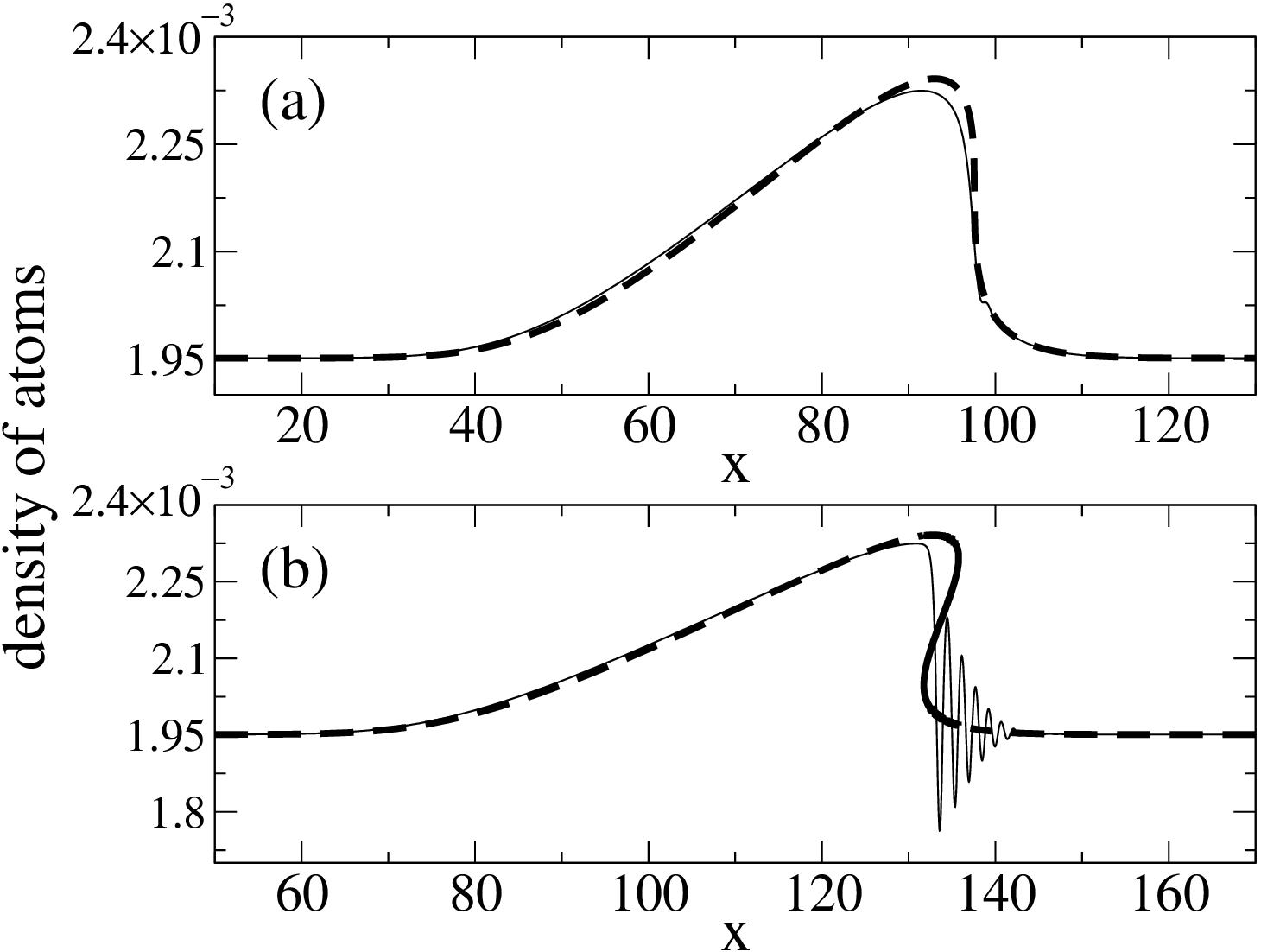}
\includegraphics[scale=0.3,clip=true]{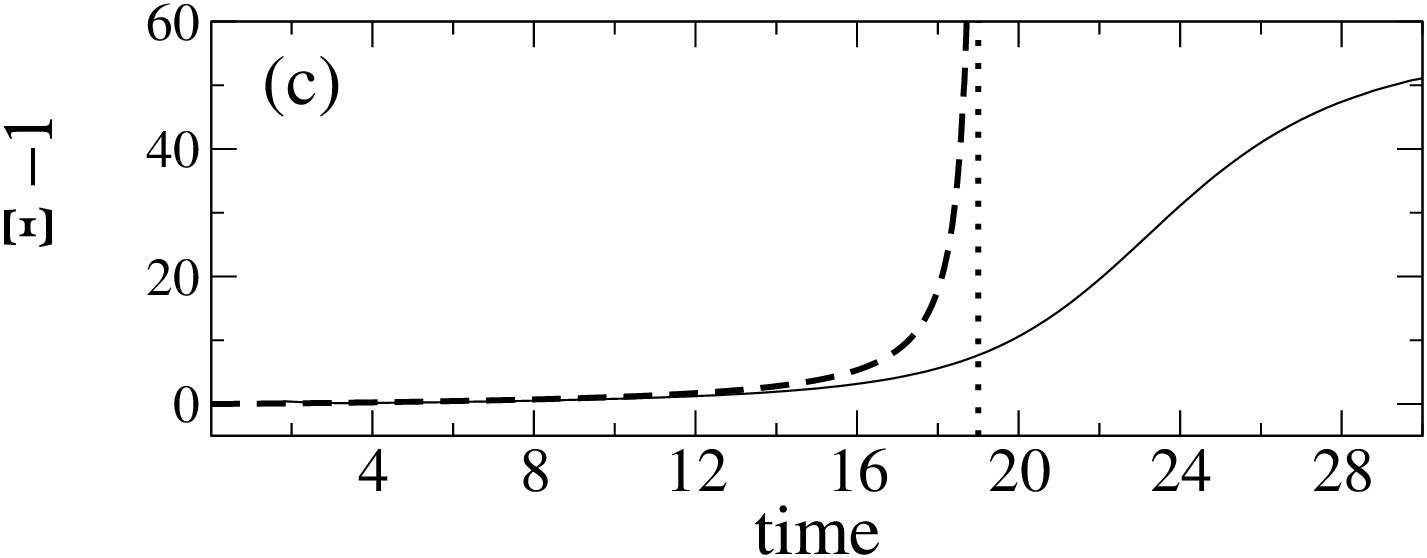}
\caption{Plots (a) and (b): 
density profile of the right-moving wave packet.
Dashed line -- the analytical solution (\ref{an1}), solid line -- a 
solution of
(\ref{h}) with the QP term, thick solid line in (b) shows multi-valued part
of the solution (\ref{an1}).
Plot (a) [(b)] is for $t=18.9$ [$t=27$].
Plot (c): relative slope (\ref{slope}) 
of the  right-moving wave packet. Dotted line indicates 
 time of shock formation (\ref{an4})-(\ref{an5}).
Parameters: $\eta=0.2$, $\sigma=12.5$, $t_s=19$, $x_s=93.5$, $g=7.5\cdot10^3$, $l=250$,
$\xi(2\cdot10^{-3})=0.18$. 
}
\label{large}
\end{figure}

For numerical purposes, we have fixed the dimensionless 
coupling constant $g$ to $7.5\cdot10^3$. 
In a typical 3D configuration such a value
corresponds to roughly $10^5\sim10^6$ atoms of either 
$^{87}$Rb
or
$^{23}$Na. In the present calculations such a value of $g$ guarantees that we
are well  in the Thomas-Fermi regime, which is important for  discussed
phenomena. The same qualitative results were obtained for
$g$ in the range $2\cdot10^3-5\cdot10^4$.

We have performed numerically time evolutions and have observed 
density wave packets until they reached  box boundaries being at $x=\pm l$. 
These results can be divided 
into two groups: the case of $x_s>l$ and that of $x_s<l$. In the first
situation
we do not observe shocks, although a shock-like deformation of an
impulse's shape
can be clearly visible. The agreement between analytical solution (\ref{an1})
and full numerical simulation is very good -- Fig. \ref{medium}. 
When $x_s<l$ shocks  appear. A typical situation is depicted in 
Fig. \ref{large}a-c. 

First of all, as a relative height of a wave packet grows, 
amplitude of separated density perturbations becomes a little bit smaller than
 half-amplitude of an initial perturbation. It leads to a small discrepancy
between analytical predictions and numerics (Fig. \ref{large}a). 

Second, a slope of the impulse does not become infinite at $t=t_s$, but
significantly increases. To express it quantitatively  we have defined a
relative slope by
\begin{equation}
\label{slope}
\Xi(t) = \frac{{\rm max}(-\frac{\partial\rho(x,t)}{\partial
x})}{\frac{\eta\rho_0}{\sigma} e^{-1/2}},
\end{equation}
where the denominator is  
${\rm max}(-\frac{\partial\rho}{\partial x})$ for a single wave packet 
which  density is $\rho_0+\eta\rho_0\exp(-x^2/2\sigma^2)$. After the
splitting $\Xi\approx1$ and than grows. 
As depicted in Fig. \ref{large}c the relative slope can be as large as few
tens showing  that a shock formation occurs on  a time scale $t\sim
t_s$.

Third, for $t>t_s$
density oscillations in the mean-field approach with the QP term
show up instead of a true shock 
wave front. 
As shown in Fig. \ref{large}b, for $t>t_s$ the analytical
solution  (\ref{an1})
still fits to numerics, but in a back edge of an  impulse 
{\it only}. In the front part the solution (\ref{an1}) becomes multi-valued, 
as expected from
the theory of nonlinear equations 
\cite{whitham,drazin}. We have checked that the
appearance of density oscillations is a result of coming into play of
the QP term. It is not an artifact of inability of our numerical
methods to evolve a truly shock wave front if such would appear
\cite{numerics}. Qualitatively the same results were obtained by other means
in \cite{kraenkel,ruski,kamchatnov}.

Similar density oscillations have  been recently observed
in a paper that describes shock wave formation in a Fermi (Tonks) cloud \cite{bodzio}.
We have shown there that they are an artifact of
the usage of the mean-field approximation. Indeed, 
they were absent in exact many-body calculations.
In a Fermi (Tonks) cloud there were three stages of shock dynamics:
a shock wave formation, propagation of a shock-like impulse roughly without
change of shape, and impulse's explosion leading to broadening  of density
profile. The mean-field approach has reproduced only the first stage. The
second stage was absent --  density oscillations  have appeared in
the same way as here. 
As a result, we
speculate, that it is rather doubtful that predictions of the Gross-Pitaevskii
equation at the front edge are reliable after appearance of density modulations. 

It is difficult to 
make any  definite statements about applicability of the
Gross-Pitaevskii equation for description of shocks 
without the insight into a many-body theory of interacting bosons. 
To understand the possible source of problems 
 we recall
that the N-body bosonic wave function has a general form:
$$\sqrt{\lambda_0}\phi(x_1)\cdots\phi(x_N)+ {\rm an \ orthogonal \ part}.$$
The condensate fraction  $\lambda_0\in(0,1]$ is
divided by $N$ the highest eigenvalue of a single particle density
matrix. $\phi(x)$ is the eigenvector of that matrix to 
eigenvalue $\lambda_0$. 
In the mean-field approach we assume 
$\lambda_0\approx1$. 
Nevertheless, $\lambda_0$  can change 
during time evolution. In principle, it can be noticeably 
lower than one after a shock
creation. It means that  contribution of states orthogonal to a condensate
mode, neglected in a mean-field treatment, might be significant. 
These states may smooth density 
changes by filling the oscillatory  region with atoms depleted from a condensate. 
Similar situation  
one encounters for dark soliton in a BEC \cite{jack}, namely a soliton's notch, 
which size is also of the order of the healing length, fills with depleted atoms.

\section{1D harmonically trapped systems}
\label{3}

Now we would like to find out what happens if an external harmonic
trapping  potential is
imposed on a 1D bosonic cloud. Atoms are initially placed in
both  harmonic and  laser  potentials  \cite{units} 
$$
V=\frac{x^2}{2}+ u_0\exp(-x^2/2\sigma^2).
$$
The atomic density, in the TF limit, takes form
\begin{equation}
\label{dens01}
\rho(x,0)=\frac{\mu}{g}-\frac{x^2}{2g}-\frac{u_0}{g}e^{-x^2/2\sigma^2},
\end{equation}
for $|x|\le R_{TF}=\sqrt{2\mu}$ and zero for $|x|>R_{TF}$.
Similarly as before,  relative amplitude $2\eta$ of an initial 
perturbation is defined by relation $2\eta\mu/g=-u_0/g$. We have restricted our
numerical simulations to bright perturbations in this section: $\eta>0$.
We have also assumed that the width of a laser beam 
 is much smaller  than spatial extent of the cloud. 
Under these conditions
the chemical potential $\mu$, found from the 
normalization of (\ref{dens01}),
equals 
\begin{equation}
\label{mu}
\mu(u_0)=[(1+u_0\sqrt{2\pi}\sigma/g)3g/4\sqrt{2}]^{2/3}.
\end{equation} 
Once again, we are interested in  dynamics of Gaussian-like wave 
packets created after an abrupt laser turn off. 

\begin{figure}
\centering
\includegraphics[scale=0.3,clip=true]{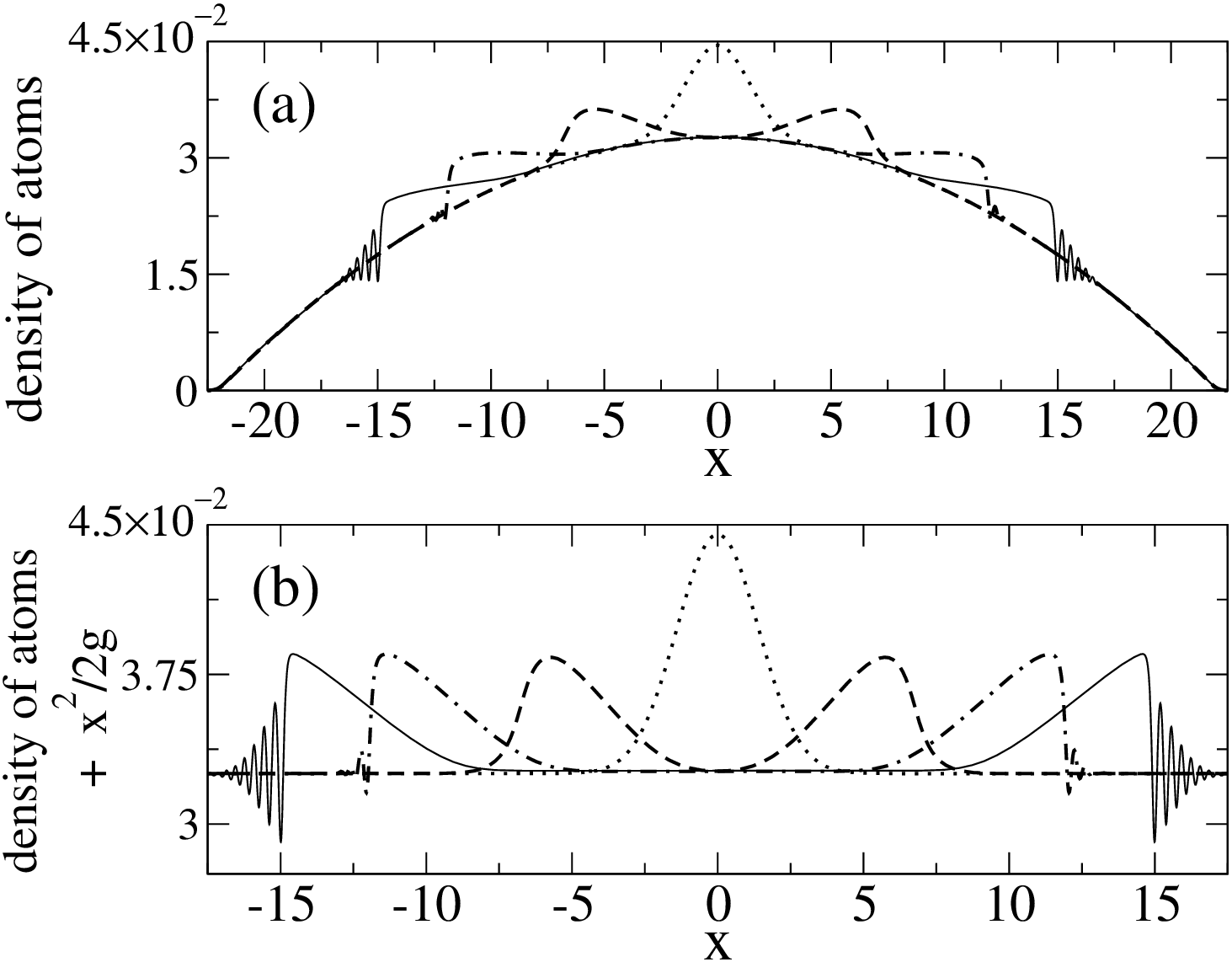}
\caption{Plot (a): density of atoms at different time moments.
Plot (b): the same as in the plot (a) but with the $x^2/2g$ term added to the
density profile in order to remove a trivial position dependence of atomic
density due to external harmonic trapping potential -- see Eq. (\ref{dens01}).
Dotted line -- $t=0$, dashed line -- $t=0.3$,
dashed-dotted -- $t=0.6$, solid line -- $t=0.8$. 
Both $t$ and $x$ are expressed 
in harmonic oscillator units \cite{units}.
Other parameters: $u_0=-90$, $\eta=0.184$, $g=7.5\cdot10^3$, $\sigma=1.4$, $R_{TF}\approx 22$.
}
\label{harm}
\end{figure}

First of all, a splitting
of an initial perturbation into two propagating outward pieces
takes place  regardless of an initial position of a density 
perturbation.
Secondly, there appears self-steepening of the bright (dark)
impulse's front (back) edge -- see Fig. \ref{harm}a. Finally, 
density oscillations 
appear showing that an  impulse enters a shock regime. 

\begin{figure}
\centering
\includegraphics[scale=0.3,clip=true]{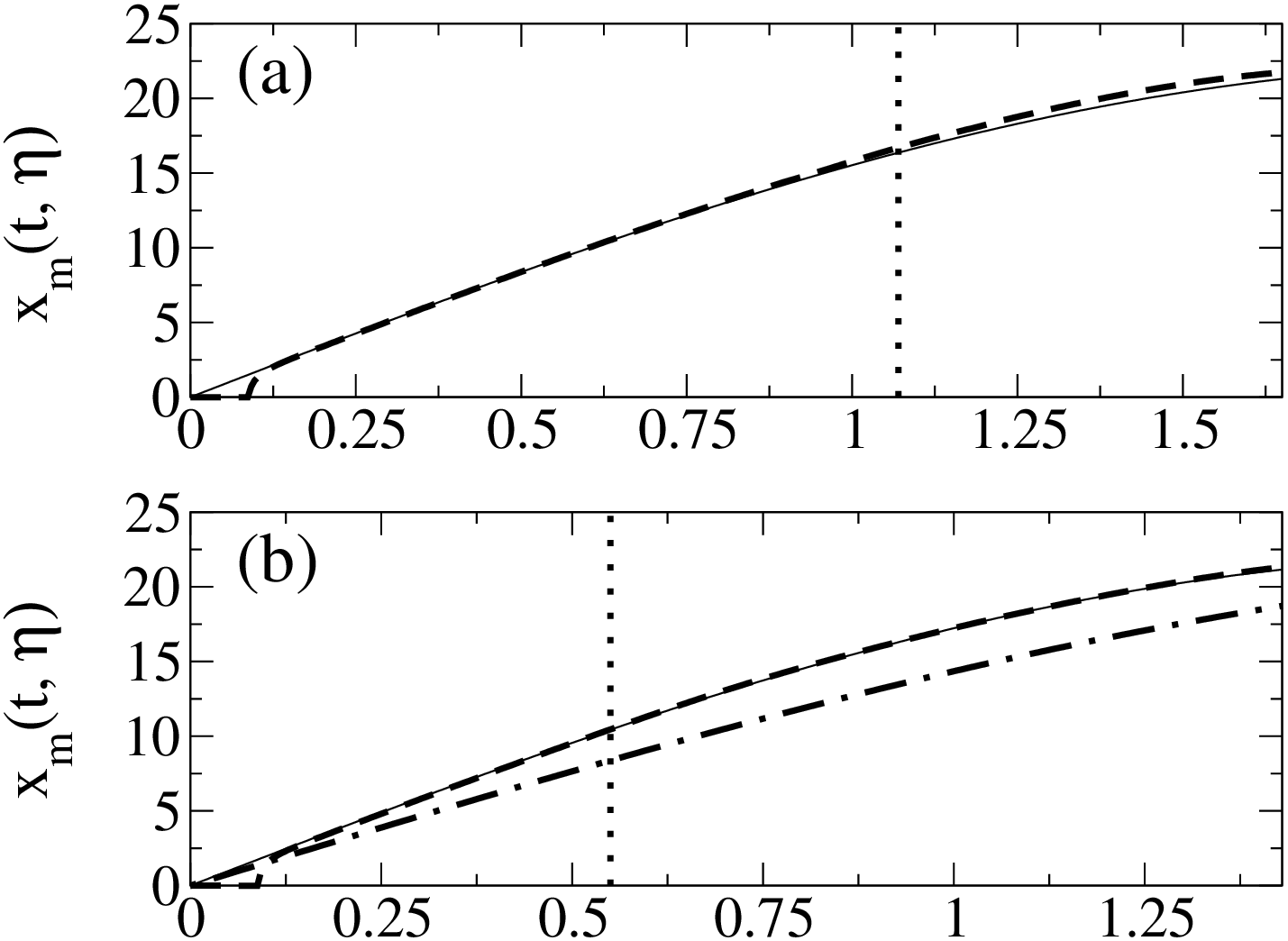}\\
\includegraphics[scale=0.3,clip=true]{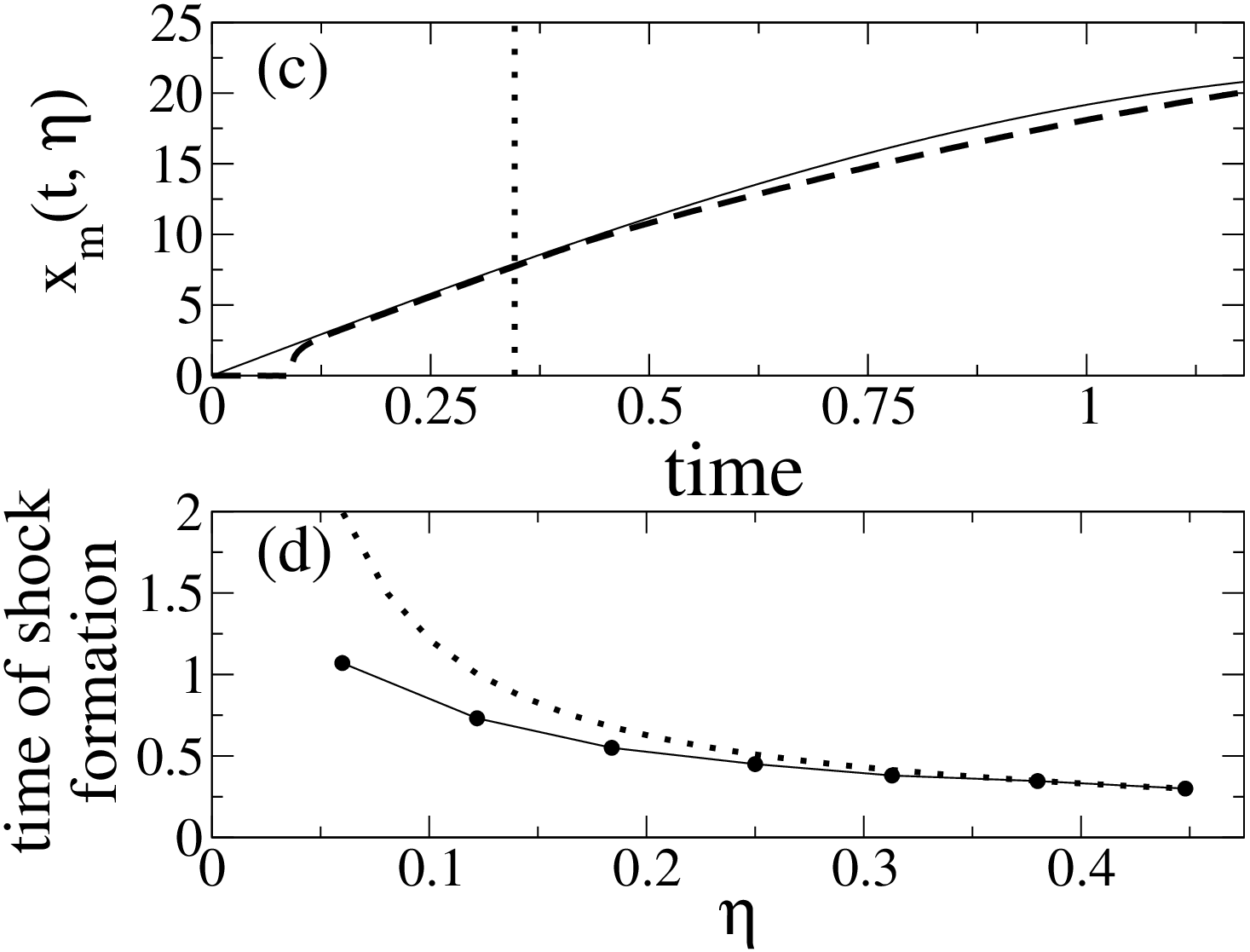}
\caption{
Plots (a)-(c): a position of impulse's maximum vs. time.
Solid line -- the semi-analytical prediction (\ref{pos2}), dashed line --
numerics on the basis of (\ref{h}) with the QP term,
dotted line -- time of shock formation extracted from numerics (see text). 
Dashed-dotted line on
the
plot (b): $x_m(t,0)$.
Parameters $(u_0,\eta)$ for plots (a)-(c): 
$(-30, 0.06)$,
$(-90, 0.184)$, 
$(-180, 0.39)$, respectively.
Plot (d): time of shock formation  
vs. relative amplitude of  an impulse. Solid line comes from numerics, while dotted
 line is an approximation (\ref{timehar}). 
Both $t$ and $x$ are expressed 
in harmonic oscillator units \cite{units}.
 Other parameters: $g=7.5\cdot10^3$, $\sigma=1.4$,
$R_{TF}\approx 22$.
}
\label{pos}
\end{figure}

We were unable to find  exact solutions of Eqs. (\ref{h}) in 
a trapped case 
even when the QP term was absent. Instead of it, we can provide 
a simple semi-analytical
prediction for a speed of density perturbations on the cloud. 
To this aim we add $x^2/2g$ 
term to the density of atoms $\rho(x,t)$. 
Such a transformation  removes a harmonic
trapping term in  (\ref{dens01}) and  propagation of impulses 
is qualitatively similar to the homogeneous case  carefully discussed  above 
(Fig. \ref{harm}b). Depending on   amplitude of 
outward moving perturbations, 
in the $\rho(x,t)+x^2/2g$ picture, impulses' amplitude 
can change a little during propagation. Therefore, it is not easy to find 
out phenomenologically 
a simple 
rule that governs  motion of Gaussian-like wave-packets in a harmonically 
trapped system.

We have tried different semi-analytical formulas  for the prediction of a 
position of a wave packet's maximum, and found that the following approach
is the most accurate. 
In a homogeneous case a speed of a density perturbation
was greater than a local sound velocity by the factor
${\cal{A}}(\eta)$ (\ref{a}), with $\eta$ being a relative impulse's
amplitude (\ref{an1}).
We have assumed that a similar law holds also now. It leads to
 the following approximate equation for the position of a
maximum of $\rho(x,t)+x^2/2g$ 
\begin{equation}
\label{pos1}
\frac{dx_m}{dt}\approx{\cal{A}}(\eta) \sqrt{\mu-x_m^2/2}.
\end{equation}
Integrating (\ref{pos1}) we have found 
\begin{equation}
\label{pos2}
x_m(t,\eta)= R_{TF} \sin\left(\frac{{\cal{A}}(\eta)\sqrt{\mu}}{R_{TF}}t\right).
\end{equation}
In the limit of $\eta\to0$ the zero-amplitude 
result of \cite{kavoulakis} is recovered. 
We have checked prediction
(\ref{pos2})  for $\eta$ as large as $0.4$ and $g=7.5\cdot10^3$. Notice that it 
corresponds to  relative amplitude of an  initial perturbation equal to $80\%$!
Numerical results are depicted in Fig. \ref{pos}a-c. 
As easily noticed,
the agreement between numerics and the formula (\ref{pos2}) is satisfactory. 
We have observed that as $x_m\to R_{TF}$, Eq. (\ref{pos2}) 
underestimates slightly a position of impulse's maximum for small $\eta$ 
and overestimates 
exact result for large $\eta$ \cite{g}.
For $\eta\approx0.18$ the agreement between numerics and 
semi-analytics  is excellent -- Fig. \ref{pos}b. Despite  differences far away 
from a trap center  
it is  important to notice that the formula (\ref{pos2})
correctly works near a trap center, where density is quite homogeneous. 
Indeed, Eq. (\ref{pos2})
reproduces nicely numerics at least for $x_m(t,\eta)\le R_{TF}/2$,
$\eta\in(0.02, 0.4)$, and other parameters as in the caption of Fig. \ref{pos}.

On the basis of (\ref{pos2}) we conclude that near a trap center,
where $R_{TF}\sin({\cal{A}}(\eta)\sqrt{\mu}t/R_{TF})\approx 
{\cal{A}}(\eta)\sqrt{\mu}t$,
the position of impulse's maximum changes linearly in time and its 
speed of propagation 
equals ${\cal{A}}(\eta)\sqrt{\mu}$. This observation is in qualitative 
agreement with the experiment  performed 
in a quasi-one-dimensional configuration \cite{kettfale}. 
The most important 
message from the formula (\ref{pos2}) is that the speed of finite size wave 
packets can differ
significantly from the zero-amplitude result by the factor ${\cal{A}}(\eta)$. 
For a realistic value of $\eta=0.3$ a speed of a wave packet's maximum
is  greater from the zero-amplitude prediction 
by more than $40\%$! We would like to stress, however, that these estimations 
are obtained for  idealized case of a  one-dimensional BEC, while the real
experiments are typically performed in quasi-one-dimensional setups. Further
studies are needless for clarification of whether similarly strong dependence
of the speed of finite size wave packets on their amplitude survives in a
quasi-one-dimensional configurations similar to that of \cite{kettfale}.
The propagation of tiny density perturbations in
harmonically trapped quasi-one-dimensional setups was discussed in 
\cite{kavoulakis,quasi}. These papers show that the  tight
confinement in cigar-shaped configurations changes the sound
velocity expected for  one-dimensional harmonically trapped 
systems by a factor $1/\sqrt{2}$.
If the same property 
would hold for finite density perturbations, the application of our results
from this section to quasi-one-dimensional systems would be straightforward.

From our simulations we have extracted time of shock formation, defined now
by a time moment when  density oscillations in a shock front have
amplitude  large enough to lower  density $\rho(x)+x^2/2g$ below the background 
value $\mu/g$. We have found that shock-like behavior shows up for
any  $\eta\in(0.06,0.45)$ and other parameters as in the 
caption of Fig. \ref{pos}. The shock behavior of a large wave packet takes
place in a relatively homogeneous part of the cloud where a sound velocity
is $\approx\sqrt{\mu}$. Replacing in the formula (\ref{tshock})
$\sqrt{g\rho_0}$ by $\sqrt{\mu}$ the following estimation for time of shock
creation can be derived
\begin{equation}
\label{timehar}
t_s\approx\frac{2\sigma}{3(\sqrt{1+\eta}-1)\sqrt{\mu}}.
\end{equation}
As depicted in Fig. \ref{pos}d, this semi-analytical formula nicely reproduces 
time of a shock formation for large enough $\eta$, and parameters 
of Fig. \ref{pos}, despite its approximate
character. The same good agreement is found for $g=10^4$ and
$g=1.25\cdot10^4$. For $g$ lower than $7.5\cdot10^3$, the case 
presented on the plot,
discrepancy between estimation  (\ref{timehar}) and exact calculation 
grows up, e.g. (\ref{timehar}) overestimates time of shock formation 
at $\eta=0.4$ and $g=2.5\cdot10^3$ by about $10\%$.

\section{Conclusions}

We have considered  propagation of broad density perturbations in a BEC.
Our theoretical approach fully accounts for  nonlinear effects
in sound propagation so it is well beyond the common  zero-amplitude 
approximation \cite{quasi,kavoulakis}, and allows for description of
shock formation.
As a result, quantitative predictions concerning propagation of  large
density perturbations in a homogeneous system were given. 
Extension of these results to harmonically trapped
problems  resulted in a semi-analytical description 
being in a satisfactory  agreement with numerics.

We have focused on  1D systems recently realized 
experimentally \cite{kettlow}, but we expect  that our work may
provide a theoretical basis for investigations of quasi-one-dimensional 
setups widely studied in laboratories.
The easiest for
experimental inspection are predictions for harmonically trapped condensates. 
We expect that  
also findings for  homogeneous bosonic clouds can be verified with the help of, e. g. 
atomic wave guides \cite{kettchip}. 

It is worth to point out that shock-like structures  have been already studied 
experimentally in a two component BEC \cite{hau}, where  
time evolution of  dark narrow density perturbations was
observed. The initial size of a wave packet in \cite{hau}
 was of the order of the healing length, which made observation of  
 impulse's self-steepening, a clear signature that shock formation takes place,
 very difficult.
We expect that the experimental setup proposed by us 
should give  more convincing results. In fact, it seems that a slight change in a
setup of
\cite{kettfale} together with a little bit higher  measurement accuracy
should verify our predictions at least qualitatively.

From the theoretical side we would like to note that 
formation of shock-like solutions in a harmonic trap  
has been noticed  in \cite{kavoulakis}. Our considerations provide
an analytical insight into their numerics. It is also worth to mention, that
 it is possible to find 
 an approximate analytical solution of hydrodynamical equations (\ref{h})
with the quantum pressure term. That solution also exhibits shock behavior, 
and its properties have been discussed in \cite{kraenkel,ruski,kamchatnov}. 

It is also interesting  to  discuss   differences between 
the traveling wave solutions presented in this paper and 
recently investigated, both theoretically and experimentally, 
solitonic  solutions  in a BEC \cite{solit}. 
Our paper  describes density perturbations
which change their shape during time evolution, while 
soliton-like density wave packets propagate in dispersionless manner \cite{drazin}.
The main difference between time evolution of wave packets considered
here and solitonic ones results mainly from the difference in their width.
The width of a soliton is of the order of the healing length, which makes the
quantum pressure term in Eq. (\ref{h}) considerable. This allows for finding 
a solution that propagate without changes of shape.
Our solution, Eq. (\ref{nonlin}), 
describes a perturbation that  is  much broader than
the healing length. In this case the quantum pressure term is negligible
and instead of solitonic solutions shock waves show up. 
It is also worth to realize that 
the initial shape of a shock-like solution, determined by the function 
$f(x)$ in  Eq. (\ref{nonlin}), 
 can be arbitrary as long as $f(x)$ changes significantly 
 on length scales much larger than the healing length. 
Conversely,  solitons   have precisely defined shape \cite{drazin}.

One of the main open questions is how the shock-like impulse's front evolves in time. 
We have pointed out that the mean-field approach  gives rather 
doubtful answer and suggested a possible explanation.
Further investigations to resolve that issue will be a subject of future
studies. 
Unfortunately, contrary to the case of Fermi 
(Tonks) gases \cite{bodzio}, it is very difficult to solve a
many-body problem exactly.

Finally, it is worth to point out that both a splitting of an initial
perturbation and a shock wave formation take place in classical
hydrodynamics \cite{somerfeld}. It suggests that qualitatively 
the same  phenomena can be observed in a 
cold bosonic (fermionic) gas  above the condensation (Fermi) temperature. 
The crossover from a classical non-degenerate  to a quantum 
degenerate regime 
can be  an interesting subject for future experimental and 
theoretical investigations.
In particular, differences in properties of cold bosons and fermions 
can be very exciting. Indeed, e. g. 
a sound velocity of 1D ultracold bosons (fermions) is $\sim\sqrt{\rho}$ 
($\sim\rho$), while in the non-degenerate regime it has to have
the same dependence on density for bosons and fermions.

I would like to acknowledge  discussions with Aneta Damska and Zbyszek Karkuszewski.
I am also grateful to Andreas Buchleitner for 
drawing my attention to  Ref. \cite{somerfeld} and for his hospitality 
during preparation of the manuscript at MPIPKS  Dresden. I want  also
to thank L. P. Pitaevskii for showing me the Ref. \cite{ruski}.
This work was mainly supported by  KBN project 2 P03B 124 22. Additional
support from the Alexander von Humboldt Foundation, Deutsche
Forschungsgemeinschaft and 
VolkswagenStiftung under the grant "Entanglement measures and the influence of
noise" is also gratefully acknowledged.


\begin{thebibliography}{99}
\bibitem{units} Let $m$ be a mass of a bosonic particle.
For a case of particles in a box we 
take arbitrary length scale $l_0$ for a unit of length and 
$ml_0^2/\hbar$ for a unit of time.
In a harmonically trapped system a unit of length equals
$\sqrt{\hbar/m\omega}$ and a unit of time is $1/\omega$ ($\omega$
is a harmonic trapping frequency).
\bibitem{sound} By a sound velocity we understand a speed of propagation of 
infinitesimally small density perturbations.
\bibitem{numerics} We use the Bulirsch-Stoer method \cite{recipy} in a grid basis 
for integration of Eqs. (\ref{h}). Such an approach can not correctly describe
shock propagation since it calculates derivatives across a shock wave front.
We have performed  numerical simulations of (\ref{h}) without and with the
QP term. In the first case the method produced results not converging
on arbitrarily large grid -- as expected from (\ref{an1})-(\ref{xs})
shocks have appeared in the numerical solution.
Conversely, the results 
of calculations with the QP term where perfectly converged 
indicating  that  shocks were absent in the 
solution of (\ref{h}) with the QP term. We have also performed the same
calculation in the appropriately truncated Fourier basis to avoid potential problems with
calculation of derivatives. As expected, the results were
the same as those obtained on a grid for  large enough size of
the Fourier basis.
\bibitem{g} The same qualitative results are obtained for
$g$ in the range $2.5\cdot10^3 - 1.25\cdot10^4$.
\end{thebibliography}
\end{document}